# Implicit learning of object geometry by reducing contact forces and increasing smoothness.

Daohang Sha, James L. Patton, Ferdinando A. Mussa-Ivaldi
*Sensory Motor Performance Program, Rehabilitation Institute of Chicago and Northwestern University Medical School*
*E-mail: d-sha, j-patton, sandro@northwestern.edu*

## Abstract

*Moving our hands smoothly is essential to execute ordinary tasks, such as carrying a glass of water without spilling. Past studies [1-3] have revealed a natural tendency to generate smooth trajectories when moving the hand from one point to another in free space. Here we provide a new perspective on movement smoothness by showing that smoothness is also enforced when the hand maintains contact with a curved surface. Maximally smooth motions over curved surfaces occur along geodesic lines that depend on fundamental features of the surface, such as its radius and center of curvature. Subjects were requested to execute movements of the hand while in contact with a virtual sphere that they could not see. We found that with practice, subjects tended to move their hand along smooth trajectories, near geodesic pathways joining start to end positions, to reduce contact forces with constrained boundary, variance of contact force, tangential velocity profile error and sum of square jerk along the time span of movement. Furthermore, after practicing movements in a region of the sphere, subjects executed near-geodesic movements, less contact forces, less contact force variance, less tangential velocity profile error and less sum of square jerk in a different region. These findings suggest that the execution of smooth movements while the hand is in contact with a surface is a means for extracting information about the surface's geometrical features.*

## 1. Introduction

When we move the hand from one point to another – for example when we reach for a glass on the table – we can chose among a variety of possible pathways. However, past studies [1] have shown that we tend to choose the shortest path for the hand and we tend to move the hand with a simple temporal profile of velocity: start from rest, reach a maximum velocity and go back to rest. This deceptively simple behavior has been interpreted as a tendency to produce maximally smooth hand movements at the expenses of complex coordination of muscles and limb segments [3]. Even people who are blind form birth exhibit this tendency [4, 5], which should not therefore be attributed to visual processing mechanisms.

Mathematically, the smoothness of a movement is captured by low or minimal values of high-order temporal derivatives of position. Flash and Hogan [3] applied the concept of smoothness to the study of hand movements by deriving the trajectory that minimizes the amplitude of the first temporal derivative of acceleration, or "jerk", given a fixed start and end position. They found that a minimum-jerk model is adequate to capture the main features of unconstrained hand movements. Hogan and Flash's minimum-jerk solution has been only applied to unconstrained movements in a "flat" space governed by the rules of Euclidean geometry. According to these rules, a straight segment is the shortest path between two points. Thus, minimum-jerk movements take place along straight segments. Euclidean geometry, however, does not apply when one is constrained to remain on a curved surface. On such surface, the minimum-distance path between two points, also called a "geodesic" path, is determined by the local curvature of the surface [6]. Over a sphere, for example, the geodesic path *joining* two points is the intersection of the sphere with the plane passing through the two points and the center of the sphere. This is how the shortest routes of airplanes and ships are calculated. Using numerical methods combined with Lagrange multipliers (see methods), we have derived the minimum jerk solution over a spherical surface [7]. The numerical computation is rather cumbersome: it involves solving a two-point boundary value problem with 18 boundary conditions. However, the solution is ultimately simple and intuitive (Figure 1 C): the minimum jerk trajectory takes place over the geodesic path joining start and end point. The tangential velocity follows a bell shape temporal profile.

An important observation follows from the theoretical result on constrained minimum jerk: if the minimum-jerk motion takes place over a geodesic segment, then executing minimum jerk or, more generally, smooth movements over a surface involves knowing fundamental geometrical features of the surface, such as its local curvature. This observation leads us to formulate the following hypothesis: *when constrained to move their hand over a curved surface, subjects develop a model of the surface structure through the progressive recovery of* *smooth (minimum-jerk) movements and the reduction of contact forces.* According to this hypothesis, smoothness of movement is not only a criterion for efficiency, but also (and perhaps, mostly) a means for learning the geometry of the constraints we come in contact with.

## 2. Methods

**Apparatus**. A PHANToM 3.0 robot (SensAble Technologies, Boston, Figure 1A) provided an object-oriented programming environment for force-feedback haptic interface [8]. The robot rendered a virtual, semi-spherical surface of radius 20 cm with a stiffness of 1 N/mm in both sides of boundary for all experiments with the update frequency of 1kHz. The surface is desired elastic surface. There is no friction along the surface, only contact force in radial direction. The computer-controlled robot will record the hand position in Cartesian coordinates in 3D space and the contact force on the hand corresponding to the position of hand in three directions with a 10ms sample interval. The computer can automatically count the number of movements and show the number on the screen by a Graphic User Interface implemented by MS VC++ for monitoring the entire experiment. Movements were separated by means of a small velocity threshold 0.005m/s. A tone after each movement provided feedback to help maintain the movement speed between 0.6 and 1.0 m/s.

A

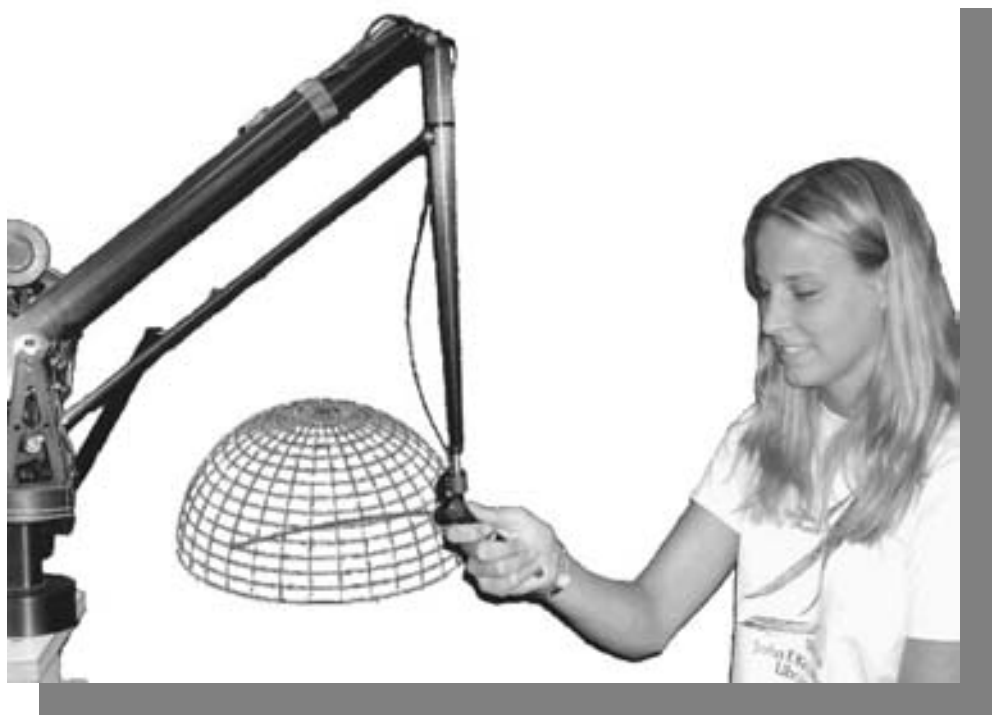

B

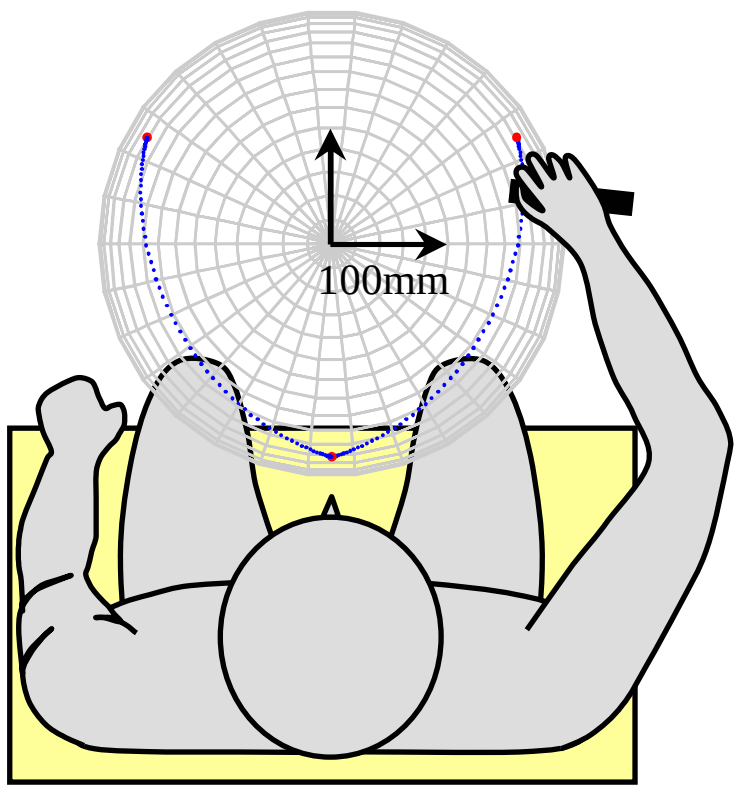

C

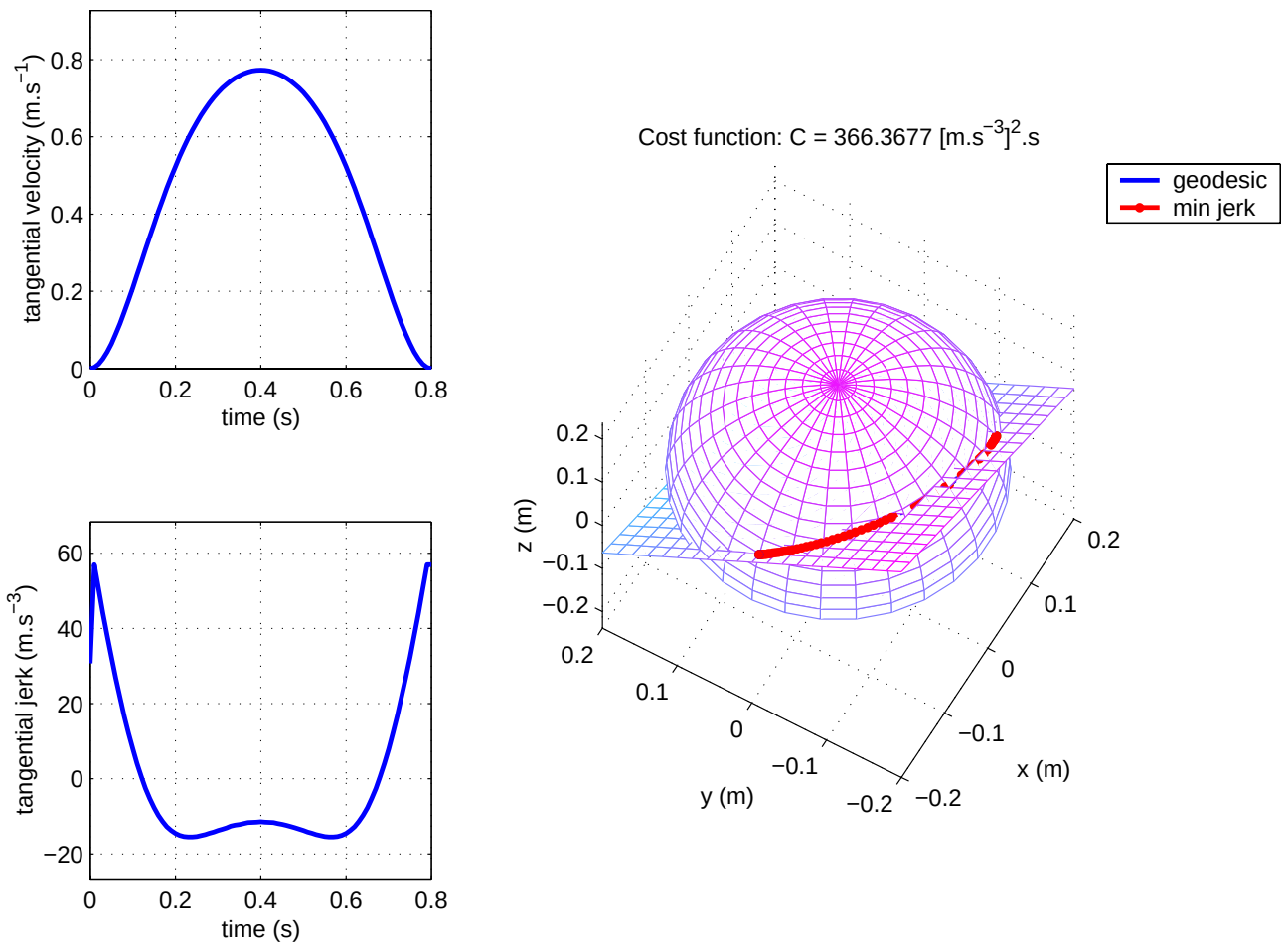

Figure **1** Configuration of experiment: **A.** Subjects were seated and held the handle of the lightweight PHANToM robotic device; **B.** There was never any mention of the constraining surface. Before the experiment, subjects were shown the three real targets on an equilateral triangle they would attempt to move to in the experiment in the right front of subject; **C.** Geodesic and numerical solution of minimum jerk movement along spherical surface. It shows that the minimum jerk solution over a sphere takes place on a geodesic segment (right). Two graphs in left column are the tangential velocity and tangential jerk of reaching movement along surface respectively. The profiles of tangential velocity and jerk are similar to the profiles of reaching movements in free space [3]

**Experiments.** A total of twenty-two right-handed, healthy subjects with no history of orthopedic or neurological disorders volunteered to participate (age 25-42; 12 male, 10 female). Before beginning the experiment, each subject signed a consent form that conformed to federal and Northwestern University guidelines. Subjects were seated and held the handle of the lightweight PhantomTM robotic device. Three targets were used in these experiments. The targets were located in front of the subjects, at the vertices of an equilateral triangle, which was contained on the horizontal plane at 80mm above the center of the constraint sphere (Figure 1B). The highest point of the sphere was situated about 30 cm below the right shoulder, centered directly in front of the subject. During the experiment, subjects could not see either the targets or their hand. They were required to memorize the locations of the three targets and then they were blindfolded. This paradigm was adopted to avoid effects induced by geometrical biases that could be induced by the visual systems. Since the experiment was not aiming at assessing the accuracy of target reaching, the simplest approach was to remove all visual inputs. There were two phases in the experiment, i.e. training (only on right side) and test (on left side). To avoid muscle fatigue, subjects were asked to rest for 20 seconds or longer every 40 movements.

**Analysis – theoretical minimum jerk.** We calculated theoretical minimum-jerk trajectories on a spherical surface by extending the optimization problem of Flash and Hogan [3, 7], defined as the movement that minimizes the time integral of the square of the rate of change of acceleration. This is a standard constrained optimization problem represented by the cost function

$$C = \int_0^{t_f} \left[ \left( \frac{d^3 x}{dt^3} \right)^2 + \left( \frac{d^3 y}{dt^3} \right)^2 + \left( \frac{d^3 z}{dt^3} \right)^2 + \lambda g(x, y, z) \right] dt, \quad (1)$$

where $\lambda$ is a Lagrange multiplier penalizing the undesired penetration into the constraining surface g, a hemisphere:

$$g(x, y, z) = x^2 + y^2 + z^2 - r^2 = 0, \quad (2)$$

where $r$ is the radius of the sphere. By using the optimal method to solve an Euler-Poisson equation [9], the resulting problem is a two-point boundary-value problem with 18 boundary conditions, where a direct, analytical solution is not possible. We used the MATLAB BVP4C function to calculate a minimum-jerk trajectory for an arbitrary surface. For the experiments of this study, the shortest path between two points on a sphere is the geodesic, formed by passing a plane through the center of the sphere, the start and the end point of the movement. Optimization results converged to the geodesic path (blue traces in Figures 1B and C), with a bell-shaped speed profile that resembled the solutions obtained on the plane.

**Analysis.** Tangential velocity was derived after fitting the sampled movement points with eight-order polynomials. Movements were separated using speed thresholds (above 0.025 m/s). The haptic service loop was updated at 1 kHz and data were sampled and stored at 100 Hz. Hypotheses were tested using signed-rank paired tests (with distribution of data is non normal) at an alpha level of 0.05. We measured two performance parameters associated with constrained motions: the average distance error and the average contact force. The average distance error of one movement was defined as the average difference between the trajectory and the corresponding ideal, minimum jerk movement divided by the length of minimum jerk movement. Average contact force (ACF) was defined as sum of distance weighted contact forces at each sample divided by the length of trajectory. Contact force variance is the variance of ACF. Velocity profile error is the area error between the actual tangential velocity profile and the corresponding minimum-jerk velocity profile. Sum of square jerk is the integral of squared jerk in the movement duration.

## 3. Results

We asked 22 subjects to execute repeated movements of the right dominant hand while holding the free extremity of a Phantom 3.0 haptic interface (Figure 1A). The haptic interface was programmed to generate a spherical surface of 20cm radius. Subjects were given no information or instruction about the surface. To suppress all influences from the visual system, they were blindfolded and asked to move their hand between three remembered targets (Figure 1B), about 30 cm apart from each other, on the corners of an equilateral triangle that intersected horizontally the spherical surface. Two sets of movements were considered in this experiment: test movements, on the left side of the triangle, and training movements, on the right side of the triangle. Figures 2 and 3 summarize the experimental results obtained from a subject, who displayed a typical learning pattern. In the initial phase of the experiment, subjects executed for one and half minutes a number of test movements (60 trials). These initial trajectories did not follow a minimum-jerk pattern and they followed distinctly non-geodesic pathways. After this initial phase, subjects were asked to execute repeated movements between the targets on the right side of the triangle. This training phase lasted about nine minutes, during which subjects executed 300 movements. The trajectories during the initial minute of the training phases were also markedly different from the corresponding geodesic motions between the same targets. However, by the end of the training phase, the trajectories had shifted toward the minimum-jerk pattern. It is evident that training led to a significant shift of movements toward the geodesic path (Figures 2B, 3Aand 3B). In addition, the velocity profiles (Figure 2C) converged toward the bell-shaped profile of the minimum-jerk velocity. At the end of the training phase, subjects repeated a set of test movements (left side of the triangle) for one and half minutes. We found that the test movements (Figure 2) were shifted toward the geodesic, minimum-jerk pathway as a consequence of training with a different set of movements. This fact constitutes a generalization of learning, which strongly supports the hypothesis that the subject developed a representation of the constraint surface, rather than merely learning how to locally improve smoothness. Some evidences for shape learning are as follows.

**Average Path Deviations (APD)** As stated above, minimum-jerk movements take place along the geodesic segment joining start and end positions. An intuitive measure of smoothness is therefore the path deviation away from this geodesic segment. To quantify the degree of similarity between the path of the hand and the geodesic path joining start and end location, we measured the average distance between the two paths. The temporal evolution over repeated trials of this average distance for the same subject is shown in Figure 3A. Similar findings apply to the majority of the subjects (Figure 3B). Over the entire subject population, there was a significant reduction of distance between hand movement and geodesic path (population average reduction 12%, p< 0.008) in the training set. There was also a significant reduction of distance between hand movement and geodesic path (population average reduction 10%, p< 0.035) in the test set. However, reduction in APD was not observed in all subjects. Over 70% of cases, subjects showed the tendency to generate hand paths closer to geodesic segments both in training and in test movements (Figure 3F).

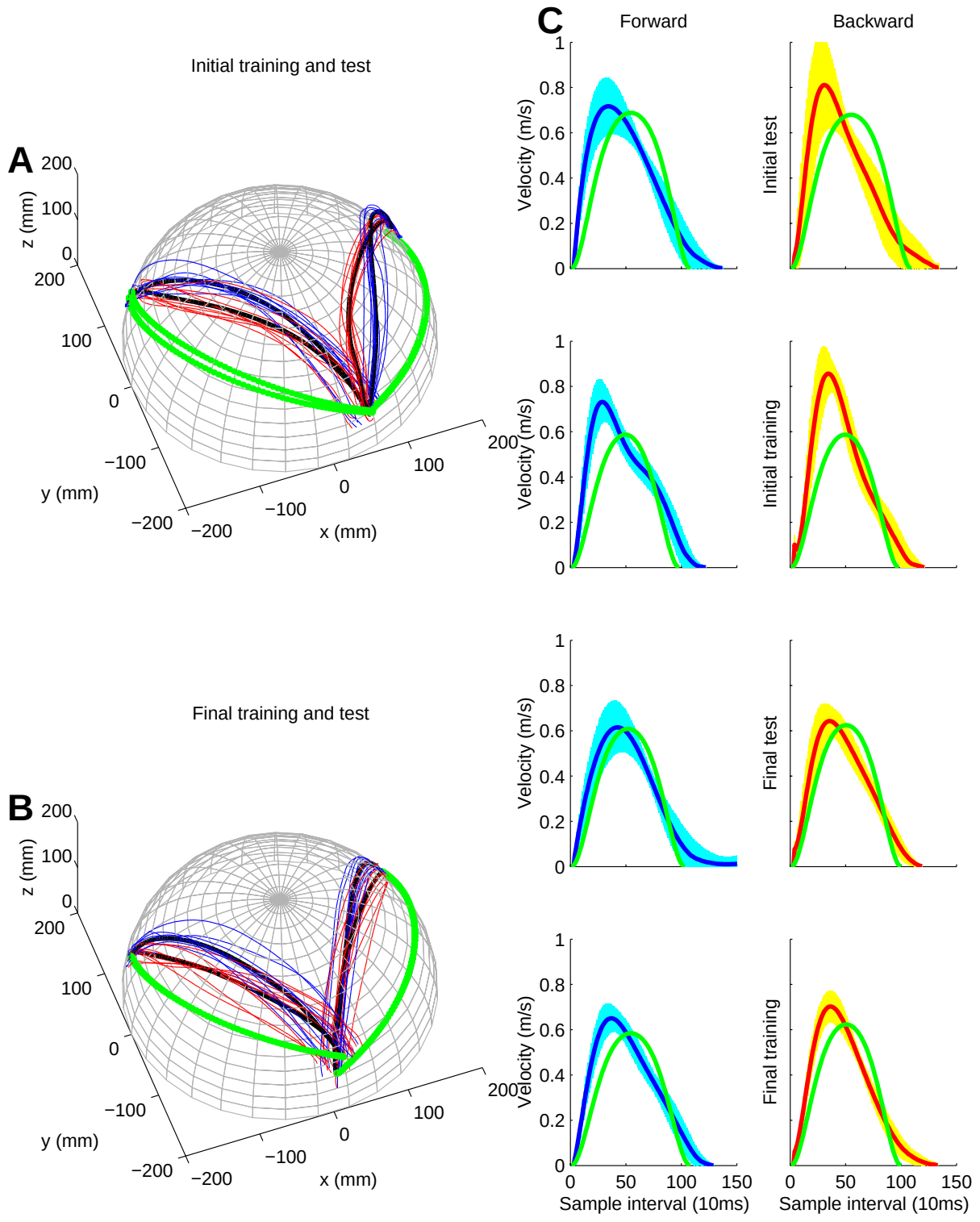

Figure **2** Trajectories and tangential velocity profiles: first column (**A, B**) is the drawing of initial and final training and test trajectories; last two columns (**C**) are the drawings of tangential velocities; solid thin lines are the actual trajectories in **A, B** and velocity profiles in **C**; blue lines represent forward movements from proximal to distal target and red lines represent backward movements (distal to proximal); solid black thick lines represent the average trajectories and velocities; solid green thick lines represent minimum jerk solutions.

**Average Contact Forces (ACF)** The reduction in contact force is an additional indication that the subject developed a representation of the spherical surface, as the amount of constant force is proportional to the degree of penetration within the boundary of the sphere. We measured the average constraint force directed toward the center of the sphere (Figure 3C). The constraint force decreased between start and end of the experiment, both for the training and for the test movements. However, the scatter plot of change in APD versus change in ACF (Figure 3E) and the related Chi square analysis indicate that the APD and ACF are statistically independent ($\chi^2 = 0.21558 < \chi^2_{0.05} = 3.8415$). This is consistent with the hypothesis that the increase of smoothness and the decrease of contact force are generated by independent processes, which both contribute to learning the surface's geometry. Over the entire subject population, there was a significant reduction of contact forces (population average reduction 29%, p< 0.00025) in the training set; there was also a significant reduction of contact forces (population average reduction 39%, p< 0.0000002) in the test set (Figure 3D). A reduction in contact forces (training and test) was observed in over 80% of cases (Figure 3F). However, only 60% of subjects reduced both path deviation and contact force.

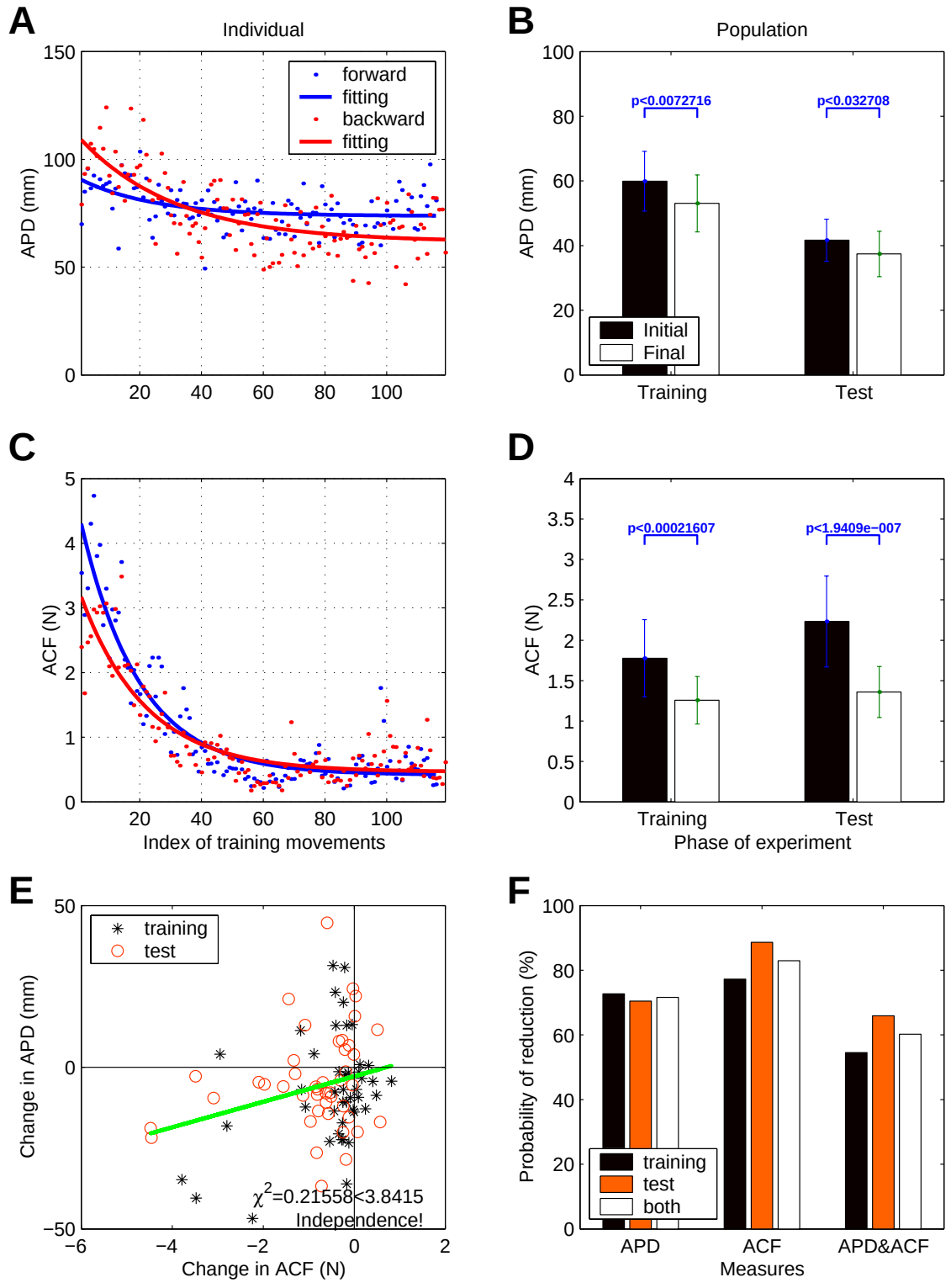

Figure **3** Average Path Deviations (**APD**) and Average Contact Force (**ACF**): **A** and **C** are the temporal evolution of APD and ACF with practice for a typical subject; all other plots show results for the entire subject population;

panels (B) and (D) show the reduction in APD and ACF, respectively, in the training and test trajectories. These changes are all significant (p<0.05); **E** is the scatter plot of APD change versus ACF change in ACF. Chi square analysis indicates that the APD and ACF are independent ($\chi^2 = 0.21558 < \chi^2_{0.05} = 3.8415$); **F** is an estimate of the probability of the reduction in APD and ACF. Over 70% of subjects reduce path deviation, over 80% of subjects reduce contact force and over 60% of subjects reduce both path deviation and contact force.

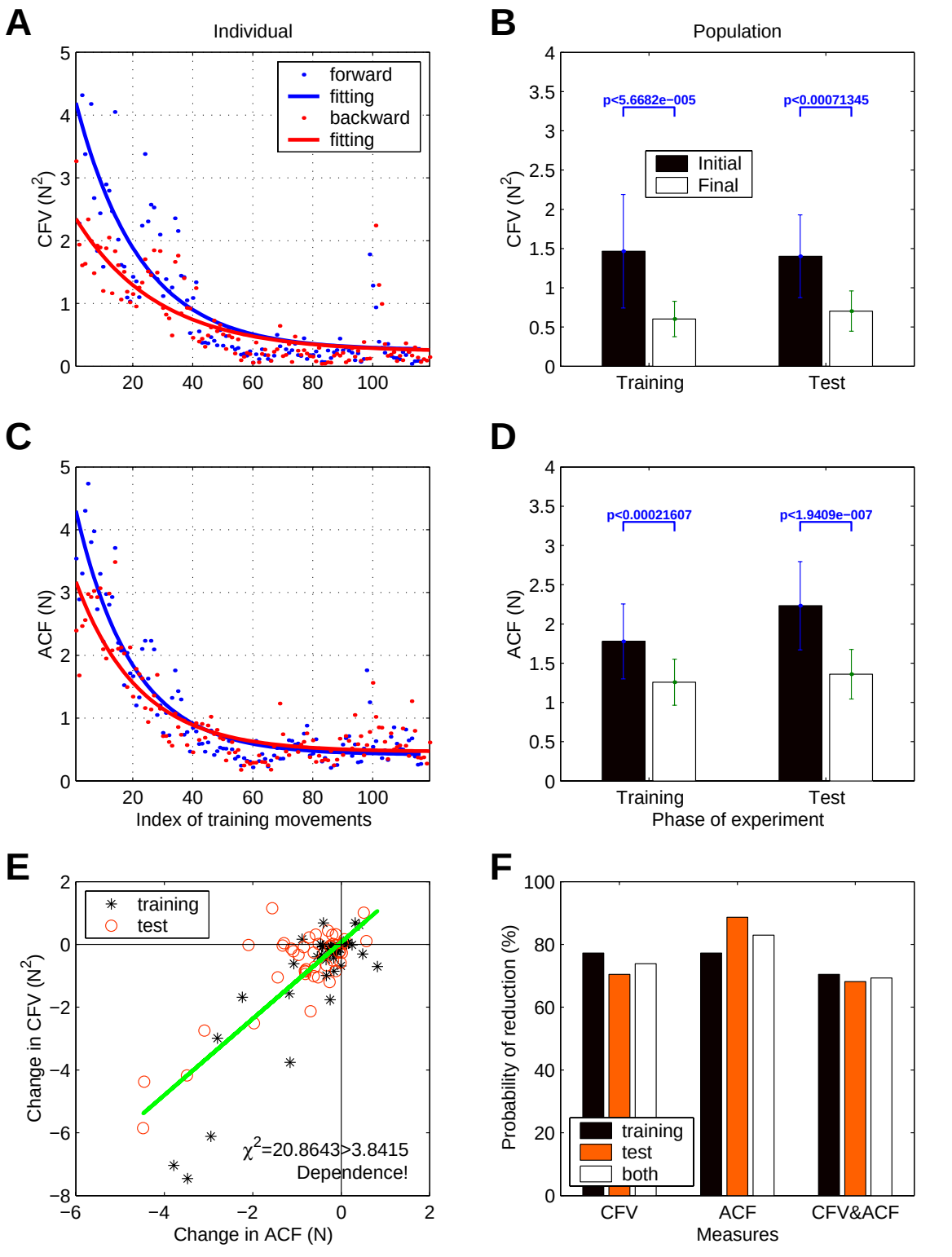

Figure **4** Contact Force Variations (**CFV**) and Average Contact Force (**ACF**): to compare CFV with ACF, ACF is plotted again here, **A** and **C** are respectively for CFV and ACF of one typical subject in the training phase; both CFV and ACF reduce for training and test (**B, D**); E shows the scatter plot of CFV vs. ACF, Chi square analysis indicate that the change in CFV and the change in ACF are dependent ($\chi^2 = 20.8643 > \chi^2_{0.05} = 3.8415$) (**E**); **F** shows that over 75% of subjects reduce CFV, over 80% of subjects reduce ACF and over 70% of subjects reduce both CFV and ACF.

**Contact Force Variance (CFV)** We also measured the variance of contact force. CFV is another measure of movement smoothness. Actually, CFV reflects the changes of penetrations along the surface. If CFV is big in size, the movement is not smooth and vice versa. It is evident that CFV decreased between start and end of the experiment (Figure 4A), both for the training and for the test movements. Over the entire subject population, there was a significant reduction of CFV (population average reduction 59%, p< 0.00006) in the training set; there was also a significant reduction of CFV (population average reduction 50%, p< 0.00007) in the test set (Figure 3B). Chi square analysis indicates that the CFV and ACF are strongly dependent ($\chi^2 = 20.8643 > \chi^2_{0.05} = 3.8415$) (Figure 4E). Over 70% of cases, this tendency corresponded to a reduction in CFV (Figure 4F). While over 65% of subjects reduce both CFV and ACF (Figure 4F).

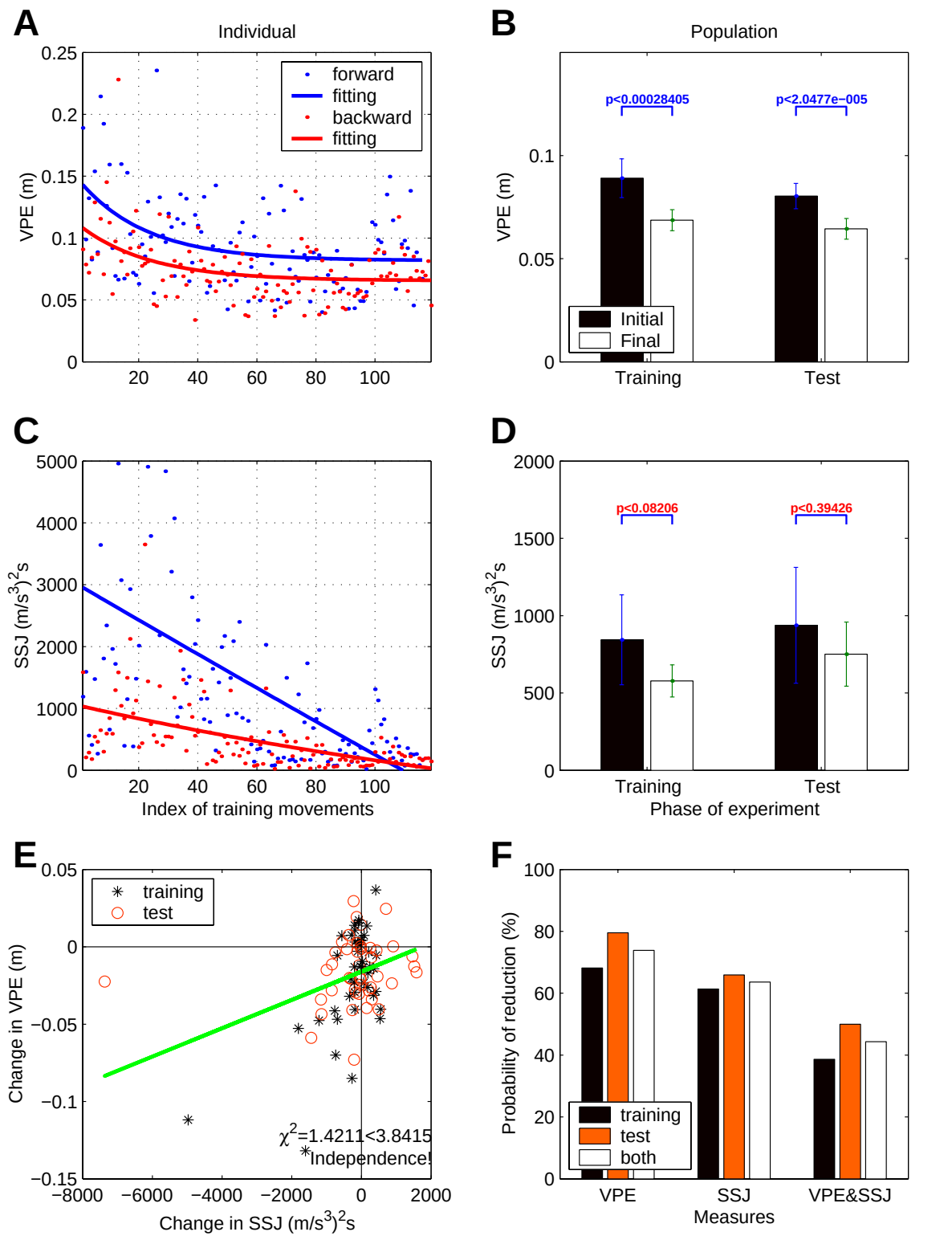

Figure **5** Velocity Profile Error (**VPE**) and Sum of Square Jerk (**SSJ**): same as displays of figure 3, **A** and **C** are the results of VPE and SSJ of one typical subject in training phase, while others are deal with the population results; **B** shows that the reductions of VPE both for training and test are significant (p<0.0003); **D** shows there is a trend in the reduction of SSJ both for training and test, but they are not significant (p>0.08); again the Chi square analysis indicates that VPE and SSJ are independent ($\chi^2 = 1.4211 < \chi^2_{0.05} = 3.8415$) (**E**); **F** indicates over 70% of subjects reduce VPE, over 60% of subjects reduce SSJ and over 40% of subjects reduce both VPE and SSJ.

**Velocity Profile Error (VPE)** VPE is the area between the actual tangential velocity and the optimal minimum-jerk velocity. PVE can be used to measure movement smoothness from another perspective. The small VPE is, the smoothest the reaching movement. VPE decreased between start and end of the experiment (Figure 5A), both for the training and for the test movements. Over the entire subject population, there was a significant reduction of VPE (population average reduction 23%, p< 0.0003) in the training set; there was also a significant reduction of VPE (population average reduction 20%, p< 0.00003) in the test set (Figure 5B). Over 70% of cases, this tendency corresponded to a reduction in CFV (Figure 5F).

**Sum of Square Jerk (SSJ)** It is evident that SSJ decreased between start and end of the experiment (Figure 4A), both for the training and for the test movements. Over the entire subject population, there was a tendency reduction of SSJ (population average reduction 31%, p= 0.1545) in the training set; there was also a tendency reduction of CFV (population average reduction 19%, p= 0.3566) in the test set (Figure 5B). Chi square analysis indicates that VPE and SSJ are independent ($\chi^2 = 1.4211 < \chi^2_{0.05} = 3.8415$) (Figure 5E). This may reflect the fact that the velocity profile and SSJ are not only dependent on the path but also on the time of movement. Over 70% of cases, this tendency corresponded to a reduction in VPE (Figure 5F). While over 42% of subjects reduce both VPE and SSJ (Figure 5F).

## 4. Discussion

The evidence described in this paper indicates that through repeated movements over a spherical surface, subjects acquire implicit knowledge of the surface geometry, as indicated by the reduction in contact force and by the increase in smoothness. This study was motivated by the observation that the execution of smooth movements over curved surfaces requires (or implies) a for of implicit knowledge about the surface geometry. Maximally smooth trajectories over the surface of a sphere take place over a segment of geodesic. The determination of a geodesic segment on a sphere involves knowledge about the center of the sphere and its radius. However, it is also possible to argue that a geodesic segment may be identified by a mechanism that enforces local smoothness, without any explicit knowledge of geometric parameters of the surface. In this case however, one may still argue that the ability to generate smooth movements over large territories of the sphere is de facto equivalent to the knowledge of such parameters,

We have begun this study by formulating the narrow hypothesis that, through repetition of reaching movements, subjects would tend to produce minimum-jerk movements, as had been observed with reaching movements in free space [2, 3]. However, the data appear to contradict this hypothesis: while we have observed a significant trend toward smoother movements over the sphere, the hand paths appear to converge toward a path that was different from the geodesic path of a minimum-jerk trajectory. While it is possible that a more prolonged exposure to the surface may lead to further reduction in jerk, the available data (e.g. Figure 3A) suggest that a plateau might have been reached by the end of the training period. However, no such plateau is evident in the sum of square jerk (Fig 5C). Our data indicate a tendency toward smoother movements, but they also suggest the possibility of a convergence to movements that are suboptimal with respect to jerk minimization. Does this suggest a different criterion?

Earlier studies have framed the study of arm motion between two alternative models: optimization of kinematics (jerk, smoothness) or optimization of a dynamic functional, such as minimum torque change [10] In kinematical models, only the geometrical and temporal properties of motion are considered, and the variables of interest are the positions (e.g. joint angles or hand Cartesian coordinates) and their corresponding velocities, accelerations and higher derivatives. Based on the observation that skilled reaching movements of the hand are smooth when described in Cartesian coordinates [1, 3], Hogan and Flash proposed that the sum of squared third derivative of Cartesain hand position, i.e. 'jerk' is minimized over the time span of movement [2, 3]. In contrast, dynamic models place the emphasis on variables such as joint torques, forces acting on the hand and muscle commands [11]. Both minimum jerk hypothesis and minimum torque-change hypothesis produce a unique solution, given the movement duration and suitable boundary conditions of the initial and final position and velocity. The minimum jerk model always predicts straight-line Cartesian hand paths with bell-shaped velocity profiles: the hand trajectory is determined only by the hand kinematics and its shape is invariant with respect to the region of the workspace. The minimum torque-change mode predicts roughly straight hand paths when the dynamics of the arm doesn't change too much during the motion, for instance, if two targets located approximately in front of the body [12, 13]. If the targets are far from each other, the hand paths of the two models are quite different, while the hand speed profiles are similar. In the minimum torque-change model, the hand path is a gently convex curve that is consistent with the empirical data for the free movement in the unconstrained horizontal plane [11].

Also, studies of more complex movements around an obstacle suggest that information about the dynamics of the arm is used in planning obstacle avoidance movements [14]. Indeed, subjects tend to select their movement paths that are perpendicular to the direction in which the arm is least sensitive when passing the near point [14]. Similarly, external movement constraints may affect kinematics of the hand. For example, when the hand moves in free space the path tends to be more curved than the path made in physical contact with a tabletop suggesting that different control strategies are involved, for constrained and free movements[15].

The observation of a trend to reducing contact forces, in parallel to the trend toward smoothness is consistent to the presence of dynamic factors that combined with kinematical planning shape the resulting hand motion. One element of caution in interpreting our data should come from the consideration that the geodesic segments are solutions for the minimum-jerk problem over a rigid curved surface, whereas the experiments were carried over a compliant virtual sphere,

While the decrease of contact force was statistically independent of the increase in smoothness, further analysis indicated a strong dependence between average contract force and contact force variations. This observation is consistent with a possible role of signal-dependent noise - characteristic of neuromuscular systems [16] in shaping the implicit learning of surface properties. Following this approach, one may speculate that the reduction of contact force may be achieved by a learning mechanism that aims at reducing fluctuations of this variable during the execution of constrained movements.

## 5. Conclusions

In summary, the repeated haptic interaction with a curved surface leads to two parallel patterns of learning: a) a trend toward smoother movements over the surface and b) a trend toward smaller interaction forces. These two trends appear to be statistically independent and both correspond to the development of a motor representation of the surface geometrical features.